\documentclass[prb,twocolumn,showpacs,amsmath,letter]{revtex4}
\usepackage{graphicx}

\newcommand{\Journal}[4]{#1 \textbf{#2}, #3 (#4)}

\begin{document}

\title{Relationship between granularity of antiferromagnet and exchange bias}

\author{Sergei Urazhdin}
\author{Phillip Tabor}
\author{Weng-Lee Lim}
\affiliation{Department of Physics, West Virginia University, Morgantown, WV 26506}


\begin{abstract}
We studied exchange bias in magnetic multilayers incorporating antiferromagent CoO doped with up to $35$ atomic percent of Pt.  The exchange bias increased with doping in epitaxial films, but did not significantly change in polycrystalline films at the lowest measured temperature of $5$~K, and decreased at higher temperatures. We explain our results by the increased granularity of the doped antiferromagnetic films, resulting in simultaneous enhancement of the uncompensated spin density and reduction of the magnetic stability of antiferromagnetic grains.

\end{abstract}

\maketitle

The asymmetry of hysteresis in ferromagnet/antiferromagnet (F/AF) bilayers, known as the exchange bias (EB) effect,~\cite{mb} has been extensively studied and utilized in magnetic devices. The EB originates from the stable uncompensated AF magnetic moments at the F/AF interface, which exert an exchange force on F approximately equivalent to an effective field $H_E$.~\cite{ohldag} The interest in the EB is stimulated by the diversity of magnetic behaviors due to a competition of several generally comparable energies: anisotropies of F and AF, volume and interface exchange energies, and the thermal energy. Understanding how the interplay of these energies affects the EB can lead to increased tunability of the magnitude, stability, and temperature dependence of the effect for device applications. 

The surface of AF films in a single-domain state is generally compensated,~\cite{remark} and thus cannot produce EB.~\cite{butler} The EB is therefore associated with a multidomain state, in which the magnetic domains are separated either by domain walls,~\cite{malozemoff} or atomically sharp boundaries. The AF can then be approximately described by a collection of single-domain grains weakly linked across their boundaries.~\cite{stiles1} Regardless of the microscopic magnetic structure of domain boundaries, their energy is generally expected to be lowered by pinning on defects and crystalline grain boundaries. In the following, we will collectively refer to the density of defects lowering the multidomain state energy as the granularity of AF.

The existence of uncompensated AF magnetic moments is facilitated by the multidomain state of AF, and therefore their density (and thus EB) should generally increase when the granularity is increased. An inverse relationship between crystalline sizes and EB has indeed been demonstrated for polycrystalline films.~\cite{takano} However, sufficiently small AF grains can become magnetically unstable due to their small anisotropy energy, in which case they do not contribute to EB.

Despite the importance of AF granularity, few experiments have directly tested its relationship with EB, and a complete picture has not yet emerged.  Simple oxide AFs are particularly attractive for such studies, because high-quality single crystals or epitaxial films can be fabricated. The effects of enhanced granularity in CoO were studied either by introducing Co vacancies,~\cite{Mgdpoing1} or by dilution with nonmagnetic Mg impurities.~\cite{Mgdpoing1,Mgdoping2} The defects were presumed to break the bonds responsible for the magnetic interactions between neighboring AF atoms. In these studies, $H_E$ increased at least for moderate doping, which was explained either by the lowering of the domain wall energies,~\cite{Mgdpoing1} or by the increased density of uncompensated interfacial AF moments.~\cite{Mgdoping2} However, the granularity introduced by the defects was superimposed on the unknown grain structure of polycrystalline CoO in one study, and a twinned structure in the other, complicating direct analysis. Additionally, these studies revealed surprisingly little evidence for decreased stability of AF grains whose sizes were effectively decreased by doping. 

We report a comparative study of EB in epitaxial and polycrystalline CoO films doped with up to $35$ atomic percent of Pt. The electronic structure of Pt is different from Co, so that its incorporation in the CoO matrix must efficiently increase granularity. We see effects of both the increased spin density and decreased AF stability. Our measurements show a significant difference between the effects of doping in epitaxial and polycrystalline films. These results are explained by the superposition of contributions from the crystalline and compositional inhomogeneities.

Multilayers with structure Co$_{1-x}$Pt$_{x}$O(6)Co(8)Cu(5)Py(5)Pt(2), where $x=0$, $0.15$, $0.25$, and $0.35$, were deposited by dc magnetron sputtering in a chamber with a base pressure of $5\times 10^{-9}$ Torr. All thicknesses are in nanometers. The top Py=$Ni_{80}Fe_{20}$ layer formed a pseudo spin valve with the exchange-biased Co, enabling electronic measurements of magnetic hysteresis via Giant Magnetoresistance (GMR).  The AF $Co_{1-x}Pt_{x}O$ was deposited by co-sputtering from Co and Pt sources in Ar/O mixture at O$_2$ partial pressure of $0.1$~mTorr, and Ar pressure of $4.6$~mTorr. The partial oxygen pressure was adjusted until films of Co and Co/Pt sputtered in Ar/O mixture exhibited negligible ferromagnetic response measured by vibrating sample magnetometry. This technique minimized overoxidation, which can lower the Neel temperature.~\cite{dieckmann}  All deposition rates were monitored by a quartz crystal oscillator. The deposition rate of CoO was kept at about $0.1$~nm/s, while the deposition rate of Pt was adjusted to the desired level. 

Polycrystalline samples were deposited on thermally oxidized Si substrates at room temperature $295$~K (RT). Epitaxial CoOPt films were deposited on polished epi-ready MgO(100) substrates at $150^o$~C after brief surface cleaning by Ar ion bombardment. Subsequent deposition of the remaining multilayer was performed in $4.3$~mTorr of purified Ar, after pumping out oxygen and cooling the sample to below $80^o$~C. The polycrystalline and epitaxial samples are labeled $px$ and $sx$, respectively, where $x$ is the Pt content. The magnetic hysteresis was measured via GMR by four-probes in Van der Pauw geometry between $5$~K and $295$~K (RT), after initial cooling in field $H=500$~Oe from RT to $5$~K.

\begin{figure}
\includegraphics[width=3.2in]{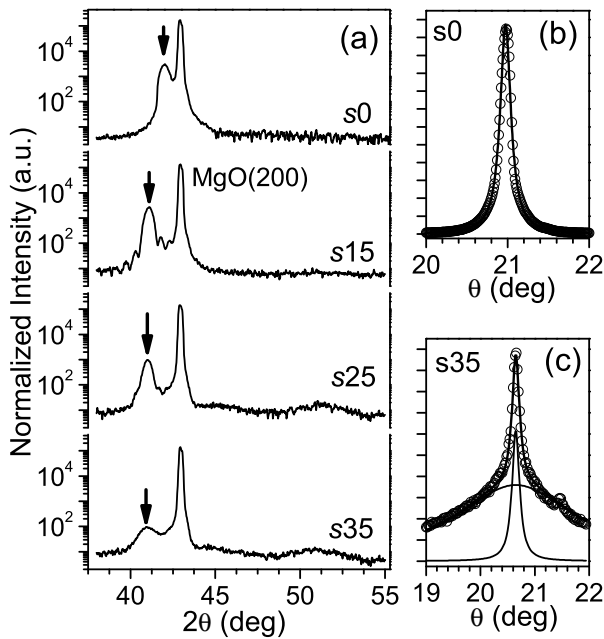}
\caption{\label{fig1} (a) $\theta-2\theta$ x-ray scans of epitaxial samples as labeled. Arrows show CoOPt(200) peaks. (b) CoO(200) rocking curve for $s0$ (symbols), fitted by a Lorenzian with FWHM of $0.15^o$ (solid curve). (c) Same for $s35$, fitted by a sum of two Lorenzians with FWHM of $0.08^o$ and $1.35^o$, as shown by three curves.}
\end{figure}

Before focusing on the magnetic properties of our samples, we first discuss the effect of Pt on crystalline properties of AF, critical to understanding the granularity of our samples. The TEM analysis of these samples has so far been unsuccessful. Instead, we have performed extensive characterization by x-ray diffraction using Cu K-$\alpha$ line. Fig.~\ref{fig1}(a) shows Bragg diffraction for the four epitaxial samples. The CoO(200) peak is at $2\theta=42.01^o$, to the left of the MgO(200) peak at $2\theta=42.93^o$. The in-plane coherence of AF films was verified by polar scans of the CoO(220) peaks (not shown). The FWHM of the rocking curve for CoO(200) was $\Delta\theta=0.15^o$ (Fig.~\ref{fig1}(b)). As the Pt doping level was increased, the CoO(200) Bragg peak first shifted to lower $2\theta$ in sample $s15$, and then decreased in amplitude without further shifts in samples $s25$ and $s35$ (Fig.~\ref{fig1}(a)).  

The correlation lengths in the direction of the scattering vector could be estimated with a Sherrer equation $\xi= 0.9*\lambda/(\cos\theta \Delta 2\theta(rad))$,~\cite{sherrer} where $\lambda$ is the wavelength of the Cu K-$\alpha$ line, and $\Delta 2\theta$ is FWHM of Bragg peak. The FWHM of the Bragg peaks were $0.5$, $0.4$, $0.4$, and $1.1$, for samples for $s0$, $s15$, $s25$, and $s35$, respectively, yielding $\xi=21$~nm, $27$~nm, $27$~nm, and $10$~nm. The first three values indicate long-range order throughout the film thickness. The values of $\xi$ for $s15$ and $s25$ are larger than the film thickness, likely due to the effect of finite-size fringes, which are especially pronounced for the sample $s15$. Their presence indicates that modest Pt doping  facilitates nearly atomic smoothness of the films. Bragg intensities for samples $s25$ and $s35$ exhibit two broad peaks around $2\theta=45.3^o$ and $2\theta=51.2^o$, which are attributed to strained Pt(200) and Pt(111) reflections, respectively. These broad peaks are likely caused by the formation of Pt clusters in films with the highest levels of doping. The rocking curves for Pt-doped samples show a superposition of a sharp peak whose amplitude decreases with doping, and a broad diffuse background increasing with doping, as illustrated in Fig.~\ref{fig1}(c) for $s35$.

The x-ray diffraction results indicate that small concentrations of Pt in $s15$ dissolve in CoO matrix. Pt clusters form at larger Pt concentrations in samples $s25$ and $s35$.  Their crystalline orientation is in registry with the CoO matrix, resulting in broad diffraction peaks in specular reflection, and diffuse background in the CoO(200) rocking curve (Fig.~\ref{fig1}(c)) due to a significant local distortion of CoO matrix around the Pt clusters. This observation will help us identify the origins of some of the magnetic behaviors described below. 

\begin{figure}
\includegraphics[width=3.3in]{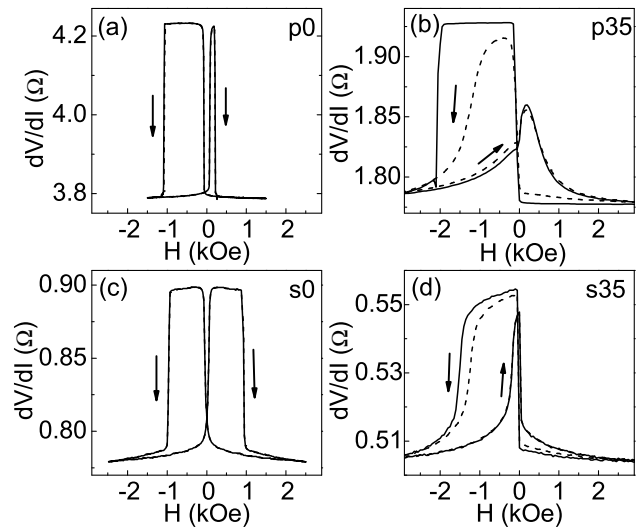}
\caption{\label{fig2} Hysteresis loops acquired immediately after cooling from RT to $5$~K in field $500$~Oe: (a) sample $p0$, (b) $p35$, (c) $s0$, (d) $s35$. Solid curves are for the first loop, and dashed curves are for the second loop. Arrows show scan directions at the reversal points of Co(8). The field is in the Co[010] direction for epitaxial samples.}
\end{figure}

Magnetic hysteresis loops were measured via GMR, after in-field cooling to $5$~K. The variations of $dV/dI$ corresponded to reversals of Co(8) and Py(5). The latter always reversed in a single sharp step at small $H=\pm (5-10)$~Oe, and thus could be easily separated from the reversal of Co(8) typically at significantly larger $H$. The field values at which resistance is halfway between the lowest and the largest value are labeled reversal fields $H^+$ and $H^-$ for increasing-$H$ and decreasing-$H$ scans, respectively. The exchange bias field $H_E$ and the coercivity $H_C$ are defined by $H_{E,C}=(H^+\pm H^-)/2$, and characterize the asymmetry and the overall width of the hysteresis loop. 

The consecutive hysteresis loops for the undoped CoO samples $s0$ and $p0$ were nearly identical, as shown in Figs.~\ref{fig2}(a),(b), where the two initial loops are indistinguishable. In contrast, Pt doping resulted in increasingly significant training effect - a decrease of the reversal field during consecutive scans.~\cite{nogues} The first two loops for samples $p35$ and $s35$ are shown in Figs.~\ref{fig2}(b) and (d), respectively. Subsequent hysteresis curves were identical to the second loop for all the samples. Fig.~\ref{fig2}(d) shows the data for $H$ parallel to the Co[010] direction. We also performed similar measurements with $H$ in Co[011] direction, yielding more square hysteresis loops, but similar values of $H_C$ and $H_E$.

The training effect originates from the instability of the initial AF configuration, which relaxes during the reversals of F due to the exchange interaction at F/AF interface.~\cite{nogues} Its presence in the Pt-doped samples is consistent with the reduced magnetic stability of smaller magnetic grains. We identify grains as the parts of the AF magnetically decoupled from the matrix due to the surrounding Pt defects and/or crystalline grain boundaries. We note that the larger training effect in polycrystalline samples indicates smaller AF grain sizes determined by a combination of crystalline grain boundaries and Pt doping. The hysteresis loop for $p35$ in panel (b) initially shows a larger asymmetry than the undoped sample $p0$ in panel (a). This behavior is consistent with the increased density of uncompensated AF moments at F/AF interface due to the increased granularity. 

Comparison of top and the bottom panels in Fig.~\ref{fig2} reveals a dramatic difference between the behaviors of polycrystalline and epitaxial samples. The undoped sample $s0$ exhibits a large $H_C$, but negligible $H_E$, consistent with the previous studies of EB in CoO(100).~\cite{CoOepitax} This result indicates negligible average uncompensated moment density as expected for single-crystalline surface of AF with low defect density. In contrast, all of the Pt-doped epitaxial samples exhibited a finite $H_E$, as illustrated for $s35$ in Fig.~\ref{fig2}(d). The appearance of EB is consistent with the increase of AF granularity, resulting in a finite uncompensated AF moment density.  

\begin{figure}
\includegraphics[width=3.3in]{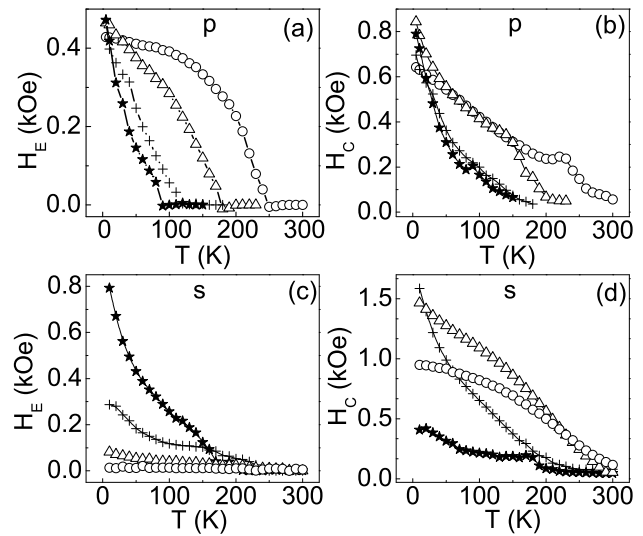}
\caption{\label{fig3} (a,b) Temperature dependence of the exchange bias field $H_{E}$ (a), and coercivity $H_{c}$ (b), for polycrystalline samples. (c,d) Same as (a,b), for epitaxial samples. Symbol types distinguish samples with different Pt concentration: circles for $x=0$, triangles for $x=15$, crosses for $x=25$, and stars for $x=35$. The field is in the Co[010] direction for epitaxial samples.}
\end{figure}

Additional information about the effect of Pt doping on the AF granularity is provided by the temperature dependence of the exchange bias and coercivity, as shown in Fig.~\ref{fig3} for all of our polycrystalline and epitaxial samples. For the following discussion, we define the blocking temperature $T_B$ of AF by the onset of finite $H_E$, which is usually correlated with a sharp increase or a bump in $H_C$. This bump is caused by torque due to the unstable AF grains, which at lower $T$ freeze out and instead contribute to $H_E$.~\cite{nogues} The data for polycrystalline samples (Figs.~\ref{fig3}(a,b)) exhibit a monotonic decrease of $T_B$ with increasing Pt doping, consistent with a decreased stability of smaller AF grains. It is also possible that Pt doping also decreases the Neel temperature of CoO, which sets the upper limit for $T_B$.  Lower values of $T_B$ lead to smaller $H_E$ over the entire measured temperature range except $T=5-30$~K, where $H_E$ becomes similar for all the samples. However, recalling that the training effect is more significant in doped samples, we conclude that doping increases the uncompensated spin density while decreasing the AF stability.

In contrast to polycrystalline samples, Pt doping of epitaxial samples results in a monotonic increase of $H_E$ over nearly the entire measured temperature range (Fig.~\ref{fig3}(c)). The opposite dependence of $H_E$ on Pt doping in epitaxial and polycrystalline samples is the main result of this paper.  The largest values of $H_E$, obtained for $s35$ at $5$~K, exceed those for polycrystalline samples by almost a factor of two. This latter result may indicate that the AF anisotropy is enhanced either due to the spin-orbit interaction at the Pt sites, or more likely due to the strain induced by incorporation of Pt atoms and clusters into the epitaxial CoO matrix. In contrast, strain is efficiently relaxed at the grain boundaries in polycrystalline samples. We note that $T_B$ is larger in epitaxial samples with the same level of doping as polycrystalline ones, which is particularly pronounced for $p35$ ($T_B=90$~K) and $s35$ ($T_B=170$~K). These differences are consistent with the larger AF grain sizes in epitaxial samples.

We also note a significantly more complex nonmonotonic dependence of $H_C$ on doping in epitaxial samples. A thorough interpretation of these behaviors will require a better general understanding of the enhanced $H_C$ in F/AF bilayers. Several possible mechanisms include torques due to the unstable AF grains,~\cite{stiles2,myFeMn} uniaxial anisotropy due to the flopping of AF moments in stable grains,~\cite{butler,stiles1} and local anisotropy due to the spatial fluctuations of the uncompensated AF moments.~\cite{stiles3,nikitenko} 
The first two mechanisms increase $H_C$ by changing the average anisotropy of ferromagnet, while the latter less explored mechanism can result in pinning of domain walls on inhomogeneities without affecting the average anisotropy.

The temperature dependence of the data shown in Figs.~\ref{fig3}(c,d) eliminates the unstable AF grains as the dominant source of enhanced $H_C$; as $T$ is lowered, such grains would become stable and begin to contribute to $H_E$, which is inconsistent with the large $H_C$ but negligible $H_E$ over a large temperature range for samples $s0$ and $s15$. One can also argue that significant local fluctuations of uncompensated AF moments are unlikely for the compensated Co(100) surface. We believe that enhanced $H_C$ in epitaxial samples is indicative of flopping of the compensated interfacial AF moments in stable grains, due to exchange interaction with F.~\cite{butler} As Pt doping is increased, the exchange interaction at the F/AF interface is increasingly accommodated by the reorientation of AF grains, likely reducing the effects of flopping.

Comparison of the top and the bottom panels in Fig.~\ref{fig3} also reveals that the dependence of $H_E$ and $H_C$ on doping in polycrystalline samples becomes weaker with increased doping level, such that the $p25$ and $p35$ data are very similar. This dependence may be expected due to the clustering of Pt in highly doped samples (Fig.~\ref{fig1}(a)), reducing the effect of doping on granularity. In contrast, the effect of increasing Pt concentration in epitaxial samples is {\it the strongest} at high doping levels (Figs.~\ref{fig3}(c,d)). This indicates that increased granularity may not be the only factor contributing to the EB in doped epitaxial samples. We believe that the magnitude of $H_E$ may be enhanced by the local  strain of the CoO matrix around Pt grains, increasing the magnetic anisotropy of AF. Such potentially complex interplay of several contributions to EB warrants more detailed quantitative simulation of this doped AF system. It would be also interesting to investigate the effect of AF thickness on the granularity and EB in this system.

In summary, we have achieved an increased granularity of epitaxial and polycrystalline CoO films by doping them with up to $35\%$ of Pt. The increased granularity resulted in a decrease of the blocking temperature and appearance of the training effect. Hysteresis measurements performed between room temperature and $5$~K showed that doping also increases the density of uncompensated AF magnetic moments, resulting in an increase of exchange bias in epitaxial samples. In polycrystalline samples, the initial increase of exchange bias was compensated by the training effect. The largest exchange bias in epitaxial samples is twice as large as in polycrystalline ones, which may be caused by an increase of the local strain-induced CoO anisotropy due to the doping. Material engineering may provide the ability to control exchange bias by doping of the host AF material to achieve an optimal combination of granularity, anisotropy, and stability. 

We thank David Lederman and Hendrik Ohldag for helpful discussions. This work was supported by the NSF Grant DMR-0747609 and the Cottrell Scholar award from the Research Corporation.


\begin{thebibliography}{99}
\bibitem{remark}  Uncompensated surfaces of real films become compensated once their finite roughness is taken into account.
\end{thebibliography}
\end{document}